\documentclass[aps,prb,a4paper,twocolumn,superscriptaddress,citeautoscript]{revtex4-1}
\usepackage{color}
\usepackage[english]{babel}
\usepackage{lmodern}
\usepackage[T1]{fontenc}
\usepackage{amsmath}
\usepackage{units}
\usepackage{grffile}
\usepackage{color}
\usepackage[bottom]{footmisc}
\usepackage{booktabs}
\usepackage{dcolumn}
\usepackage{graphicx}
\graphicspath{{figs/}}

\newcommand\hN{\hat{N}}
\newcommand{\vS}{\vec{S}}
\newcommand{\vL}{\vec{L}}

\newcommand{\J}{J}

\begin{document}
\title{Signatures of spin-orbital states of ${t_{2g}}^{2}$ system in the optical conductivity \\: The case of $R$VO$_{3}$ ($R$=Y and La)}
\author{Minjae Kim}
\email{garix.minjae.kim@gmail.com}
\affiliation{Centre de Physique Th\'eorique, \'Ecole Polytechnique, CNRS, Universit\'e Paris-Saclay, 91128 Palaiseau, France}
\affiliation{Coll\`ege de France, 11 place Marcelin Berthelot, 75005 Paris, France}

\date{\today}
\begin{abstract}
We investigate signatures of spin and orbital states of $R$VO$_{3}$ ($R$=Y and La)
in the optical conductivity using density functional theory
plus dynamical mean-field theory (DFT+DMFT).
From the assignment of multiplet state configurations to optical transitions,
the DFT+DMFT reproduces experimental temperature dependent evolutions of
optical conductivity for both YVO$_{3}$ and LaVO$_{3}$.
We also show that the optical conductivity is a useful quantity
to probe the evolution of the orbital state even in the absence of spin order.
The result provides a reference to investigate spin and orbital
states of ${t_{2g}}^{2}$ vanadate systems which is an important
issue for both fundamental physics on spin and orbital states and
applications of vanadates by means of orbital state control.

\end{abstract}
\maketitle
\section{Introduction}
\label{sec:intro}

Rare-earth vanadates, $R$VO$_{3}$ ($R$=rare-earth atom),
are one of the most intensively discussed materials for
the nature of the interplay between
the spin, orbital, and lattice
degrees of freedom.\cite{fang2004quantum,de2007orbital,khaliullin2005orbital,horsch2003dimerization,ulrich2003magnetic}
Understanding this coupling between the spin, orbital, and lattice degrees of freedom in $R$VO$_{3}$
is an important fundamental problem,
which has been a subject of debates regarding the origin of spin and orbital ordering,
electronic superexchange versus lattice distortion.
Also, this spin-orbital-lattice coupling is an important ingredient
in applications of $R$VO$_{3}$, such as in multiferroic materials\cite{hotta2007polar,jackeli2008spin,park2017charge,Varignon_electrical_control}
and solar cells\cite{assmann2013oxide,eckstein2014ultrafast,wang2015device},
by means of heterostructure engineering.
The lattice controlled spin state for $R$VO$_{3}$ is a promising route to multiferroicity,\cite{jackeli2008spin,Varignon_electrical_control}
while the spin ordering driven longer lifetime of photodoped carriers for $R$VO$_{3}$
is an important source to improve the efficiency of solar cells.\cite{eckstein2014ultrafast}

%Rare-Earth vanadates, $R$VO$_{3}$ ($R$=rare-Earth atom),
%are one of the most intensively discussed materials
%for their interplay between spin-orbital states.
%Understanding this spin-orbital-lattice coupling in $R$VO$_{3}$
%is an important fundamental problem,
%which has been a subject of debates regarding their quantum vs classical
%nature of spin-orbital states.\cite{de2007orbital,fang2004quantum,weng2010phase}
%Also, this spin-orbital-lattice coupling of $R$VO$_{3}$
%is an urgent problem for applications,
%such as multiferroic materials\cite{hotta2007polar,jackeli2008spin,park2017charge,Varignon_electrical_control} and solar cells\cite{assmann2013oxide,eckstein2014ultrafast,wang2015device},
%by means of heterostructure engineering.
%The lattice control of spin states and the spin ordering driven longer lifetime of
%photodoped carriers are important ingredients for designing
%multiferroic material and solar cells, respectively.

YVO$_{3}$ and LaVO$_{3}$ are two representative materials
of the rare-earth vanadate family which exhibit different temperature ($T$)
dependent evolutions of the crystal structure.\cite{blake2002neutron,bordet1993structural}
At high $T$, both YVO$_{3}$ and LaVO$_{3}$
have orthorhombic structures of the $Pnma$ space group with the $a^{-}a^{-}c^{+}$ type Glazer rotation
of octahedrons.
For YVO$_{3}$, with cooling, the structural transition occurs at $T$=200 K
to the monoclinic structure of the $P2_{1}/a$ space group. For lower $T$, below 77 K,
the crystal structure of this material turns into the $Pnma$ space group again.
In contrast, LaVO$_{3}$ has a larger size of the cation than that of YVO$_{3}$.\cite{blake2002neutron}
Accordingly, the $a^{-}a^{-}c^{+}$ type rotation of LaVO$_{3}$ is
smaller than that of YVO$_{3}$.
As a result, for LaVO$_{3}$, the structural
transition from $Pnma$ to $P2_{1}/a$ occurs at $T$=140 K upon cooling,
and there is no structural transition below 140 K.\cite{bordet1993structural}

%YVO$_{3}$ and LaVO$_{3}$ are two representative materials
%of $R$VO$_{3}$ which exhibit different temperature (T)
%dependent evolutions of spin-orbital states
%in relation with crystal structures.\citep{miyasaka2003spin}
%For YVO$_{3}$, with cooling,
%the structural transition firstly occurs for T=200K,
%from orthorhombic ($$$$Pnma$$$$) to monoclinic (P2$_{1}$/a).\cite{blake2002neutron,miyasaka2003spin}
%For lower T, 77K, the structure of
%this material turns into the $$$$Pnma$$$$ space group again.
%For LaVO$_{3}$, which has a larger size of cation than
%that of YVO$_{3}$, the structural transition from
%$$$$Pnma$$$$ to P2$_{1}$/a occurs for 140K upon cooling.\cite{zhou2008frustrated,miyasaka2003spin}

\begin{figure}[t]
\includegraphics[width=\columnwidth]{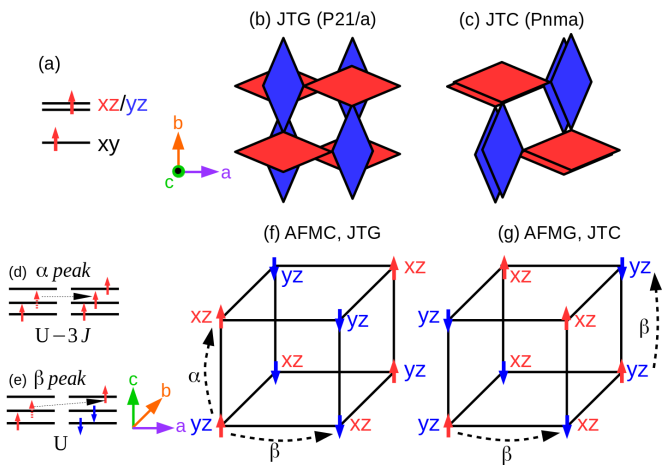}
\caption{(Color)
(a) The ${t_{2g}}^{2}$ electron configuration in $R$VO$_{3}$.
(b) and (c) G type Jahn-Teller distortion (JTG) in $P2_{1}/a$
and C type Jahn-Teller distortion (JTC) in $Pnma$, respectively.
Blue and red rhombus indicate distorted octahedrons.
(d) and (e) Multiplet configurations of optical transitions for
$\alpha$ peak (optical gap : $U-3J$) and $\beta$ peak (optical gap : $U$)
in (f) and (g) (see the text).
(f) Spin-orbital configuration in the JTG and C type antiferromagnetic
order (AFMC).
(g) Spin-orbital configuration in the JTC and G type AFM order (AFMG).
\label{fig:Introduction}
}
\end{figure}

The spin and orbital ordering of YVO$_{3}$ and LaVO$_{3}$ is related to
the crystal structure for the given $T$.
Figure~\ref{fig:Introduction}(a) shows the electronic configuration
of these materials that have two electrons in $t_{2g}$ orbitals
with the Hund's coupling induced high spin state.
The $xy$ orbital has an occupancy close to one because
of its lower energy level with respect to that of $xz$ and $yz$ orbitals,
which is driven by the rotation and distortion of octahedron.
Another electron has an orbital degrees of freedom between $xz$ and $yz$.
Fig.\ref{fig:Introduction}(b and c) shows dominant Jahn-Teller distortion 
patterns for $P2_{1}/a$ and $Pnma$ structures.
$P2_{1}/a$ structure has G type Jahn-Teller distortion (JTG), 
and $Pnma$ structure has C type Jahn-Teller distortion (JTC). 
These patterns of JT distortion lift the degeneracy on $xz$ and $yz$ orbitals
inducing G type orbital ordering (OOG) for $P2_{1}/a$ structure (JTG) and
C type orbital ordering (OOC) for $Pnma$ structure (JTC).
Fig.\ref{fig:Introduction}(f and g) shows this structure driven 
spin and orbital ordering pattern for OOG (antiferro OO in all axis) 
and OOC (ferro OO in the $c$ axis and antiferro OO in the $ab$ plane).  
With the nominal configuration in Fig.\ref{fig:Introduction}(a), 
the single electron in the $xy$ orbital induces
the antiferromagnetic (AFM) superexchange interaction between nearest neighbor atoms
in the $ab$ plane.
The OOG induces ferromagnetic (FM) superexchange interaction along the $c$ axis from
the Hund's coupling induced lower energy superexchange path along the $c$ axis.
And, the OOC induces AFM superexchange interaction along the $c$ axis, because
of that the ferro OO along the $c$ axis induces the superexchange
energy gain in the case of AFM order along the $c$ axis.
Accordingly, JTG ($P2_{1}/a$) and JTC ($Pnma$) distortions induce
C type AFM (AFMC) and G type AFM (AFMG), respectively,
in agreement with Goodenough-Kanamori-Anderson rules (GKA rules, Ref.\onlinecite{kanamori1959superexchange,goodenough1955theory,anderson1950antiferromagnetism}).
These $T$-dependent spin orderings of YVO$_{3}$ and LaVO$_{3}$ are
unambiguously determined by neutron diffraction experiments.\cite{blake2002neutron,zhou2008frustrated}

%%%%%%%%%%%%%%%%%%%%%%%%%%%%%%%%%%%%%%%%%%%%%%%%%%%%%%%%%%%%%%%%
%%%%%%%%%%%%%%%%%%%%%%%%%%%%%%%%%%%%%%%%%%%%%%%%%%%%%%%%%%%%%%%%
%%%%%%%%%%%%%%%%%%%%%%%%%%%%%%%%%%%%%%%%%%%%%%%%%%%%%%%%%%%%%%%%
\begin{figure}[t]
\includegraphics[width=\columnwidth]{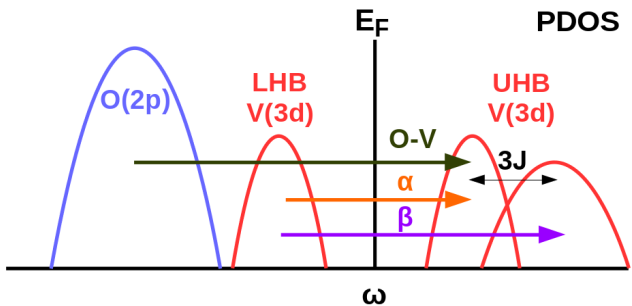}
\caption{(Color)
Schematic partial density of states (PDOS) of $R$VO$_{3}$ and
its interpretation for optical transitions, $\alpha$, $\beta$,
and O-V (O(2$p$) to V(3$d$)). LHB and UHB present
lower and upper Hubbard bands, respectively. O(2$p$) bands
contributions are also plotted.
E$_{F}$ indicates the Fermi level.
\label{fig:PDOS}
}
\end{figure}
%%%%%%%%%%%%%%%%%%%%%%%%%%%%%%%%%%%%%%%%%%%%%%%%%%%%%%%%%%%%%%%%

%Overall, the spin ordering of YVO$_{3}$ and LaVO$_{3}$ follows
%the crystal structure at given T.
%The electronic configuration of these materials is $d^{2}$
%having two electrons in the $t_{2g}$ orbitals.
%The $xy$ orbital has an occupancy close to one.
%And, another electron is shared by $xz$ and $yz$ having
%parallel spin with the electron at $xy$ by Hund's coupling.
%For the $$$$Pnma$$$$ structure, the dominant Jahn-Teller distortion is
%C type (JTC).
%Accordingly, in agreement with Goodenough-Kanamori-Anderson rules (GKA rules, Ref.\cite{kanamori1959superexchange,goodenough1955theory,anderson1950antiferromagnetism}),
%the G type antiferromagnetic order (AFMG) emerges
%for low T of YVO$_{3}$ (below 77K) as shown in Fig.\ref{fig:Introduction}(b).
%For the P2$_{1}$/a structure, the dominant JT distortion
%is G type (JTG).
%As a result, the C type AFM (AFMC) emerges
%for intermediate T of YVO$_{3}$ (77K-116K) and low T of LaVO$_{3}$ (below 140K)
%in agreement with GKA rules,
%as shown in Fig.\ref{fig:Introduction}(a).
%(See Fig.\ref{fig:OPT_YVO}(e) and Fig.\ref{fig:OPT_LVO}(e) for details)
%These spin orderings and their evolutions in YVO$_{3}$ and LaVO$_{3}$
%have been confirmed by neutron diffraction studies.\cite{blake2002neutron,zhou2008frustrated}

For the explanation of experimentally confirmed spin orderings for the given $T$,
there have been debates on orbital states for $R$VO$_{3}$.
Due to the small crystal-field inherent to $t_{2g}$ orbitals,
the quantum orbital fluctuation of $xz/yz$ could be relevant for
the $T$ regime of spin ordering,
even in the presence of pictorial JT distortions in Fig.\ref{fig:Introduction}(b and c).
Based on this small crystal-field, the resonant valance bond (RVB)
state of $xz$ and $yz$ orbitals is proposed as an orbital state of $R$VO$_{3}$.\cite{khaliullin2005orbital,horsch2003dimerization,ulrich2003magnetic}
In this orbital RVB, $xz$ and $yz$ orbitals of nearest neighbor atoms on $c$ axis
have a singlet state, and this singlet makes the RVB along the $c$ axis of the lattice.
Accordingly, in agreement with the GKA rules, the FM superexchange interaction
emerges in the $c$ axis and AFMC emerges.
In the $P2_{1}/a$ structure, both the JTG and the orbital RVB induce AFMC.
However, in LaVO$_{3}$ at $T\sim$150 K, which has the $Pnma$ structure,
the JTC in the structure induces the AFMG order\cite{zhou2008frustrated},
which is different from the orbital RVB induced AFMC order.\cite{khaliullin2005orbital}
Both view points explain experiments such as the magnon spectrum of YVO$_{3}$.
\cite{fang2004quantum,horsch2003dimerization,ulrich2003magnetic}
But, for LaVO$_{3}$,
the continuous $T$-dependent evolution of spin and orbital states
for $T\sim$150 K still remains to be resolved.
The fingerprint of spin-orbital states of LaVO$_{3}$
in the optical conductivity would resolve this puzzle, 
complementary to thermal properties such as electronic contributions to entropy.\cite{zhou2008frustrated}

%Spin-states of YVO$_{3}$ and LaVO$_{3}$ are unambiguously
%determined by neutron diffraction.
%However, there have been debates on orbital-states.
%Due to the small crystal-field inherent to the $t_{2g}$ orbital,
%the superexchange driven quantum orbital order (OO)
%could compete with the JT distortion induced classical OO,
%for both YVO$_{3}$ and LaVO$_{3}$.\cite{khaliullin2005orbital,zhou2008frustrated}
%While the classical OO supports G type OO for P2$_{1}$/a structure,
%the superexchange driven quantum OO supports the resonant valance bond (RVB) state
%of $xz$ and $yz$ orbital along $c$ axis.
%In the P2$_{1}$/a structure, both the classical OO and the quantum RVB induce AFMC.
%For LaVO$_{3}$ with T around 150K ($$$$Pnma$$$$ structure), the classical OO supports
%C type OO, and accordingly, the AFMG order emerges.\cite{zhou2008frustrated}
%On the other hand, the superexchange driven RVB of orbital supports the AFMC order.\cite{khaliullin2005orbital}
%Both view points explain experiments such as the magnon spectrum of YVO$_{3}$.
%\cite{fang2004quantum,horsch2003dimerization,ulrich2003magnetic}
%However, for LaVO$_{3}$,
%the continuous T dependent evolution of spin-orbital states
%around 150K is still remained to be resolved.

Optical conductivity is a powerful tool to probe phase transitions
in strongly correlated systems.\cite{basov2011electrodynamics}
Fingerprints of multiplet states of the correlated shell in the
optical conductivity provide information about the spin and orbital ordering.
In $R$VO$_{3}$, there have been issues on the explanation of $T$-dependent evolution of optical
conductivities in experiments 
in terms of spin and orbital states.\cite{Reul_PRB_2012,Miyasaka_JSPJ_2002,fang2003anisotropic}
However, there is an ambiguity
in the quantitative assignment of optical peaks to multiplet states,\cite{Reul_PRB_2012,Miyasaka_JSPJ_2002,fang2003anisotropic}
which we explain in the Sec.\ref{subsec:Multiplet}.
In addition, the $T$-dependent evolution of optical conductivity in the theory
to be compared with experiment is still lacking.

%Optical conductivity is a powerful probe of phase transitions
%in strongly correlated systems.\cite{basov2011electrodynamics}
%Fingerprints of multiplet states of the correlated shell in the
%optical conductivity provide information on the
%magnitude of electronic correlation, valence states,
%and emergence of condensation such as spin-orbital order and
%superconductivity.\cite{basov2011electrodynamics}
%The T dependent evolution of optical conductivity provides an
%information on the nature of spin-orbital states.

In this paper, we compute $T$-dependent
evolutions of optical conductivity of $R$VO$_{3}$ ($R$=Y and La)
using density functional theory
plus dynamical mean-field theory (DFT+DMFT).\cite{DMFT_rmp_1996,kotliar2006electronic}
The approach reproduces experimental $T$-dependent evolutions
of optical conductivity for both  YVO$_{3}$ and LaVO$_{3}$.
We have shown that the $T$-dependent optical conductivity
could detect spin and orbital orderings.
This result could be achieved by the correct assignment of
optical transitions to two peaks, $\alpha$ and $\beta$,
as shown in Fig.\ref{fig:Introduction}(d and e).
Furthermore, by imposing paramagentic (PM) state
for given structures of each $T$ in the phase diagram,
we have shown the orbital polarization driven evolution
of optical conductivity in the absence of spin ordering.
This result could be used as a reference for
probing orbital states in the heterostructure of $R$VO$_{3}$.
Also, we have shown the difference in the optical
conductivity of LaVO$_{3}$ at $T\sim$150K in various magnetic states
which would be useful for the investigation on the nature of spin and orbital states
between the orbital RVB driven AFMC and the JTC driven AFMG.

%%%%%%%%%%%%%%%%%%%%%%%%%%%%%%%%%%%%%%%%%%%%%%%%%%%%%%%%%%%%%%%%
\begin{figure*}[t]
\includegraphics[width=10cm]{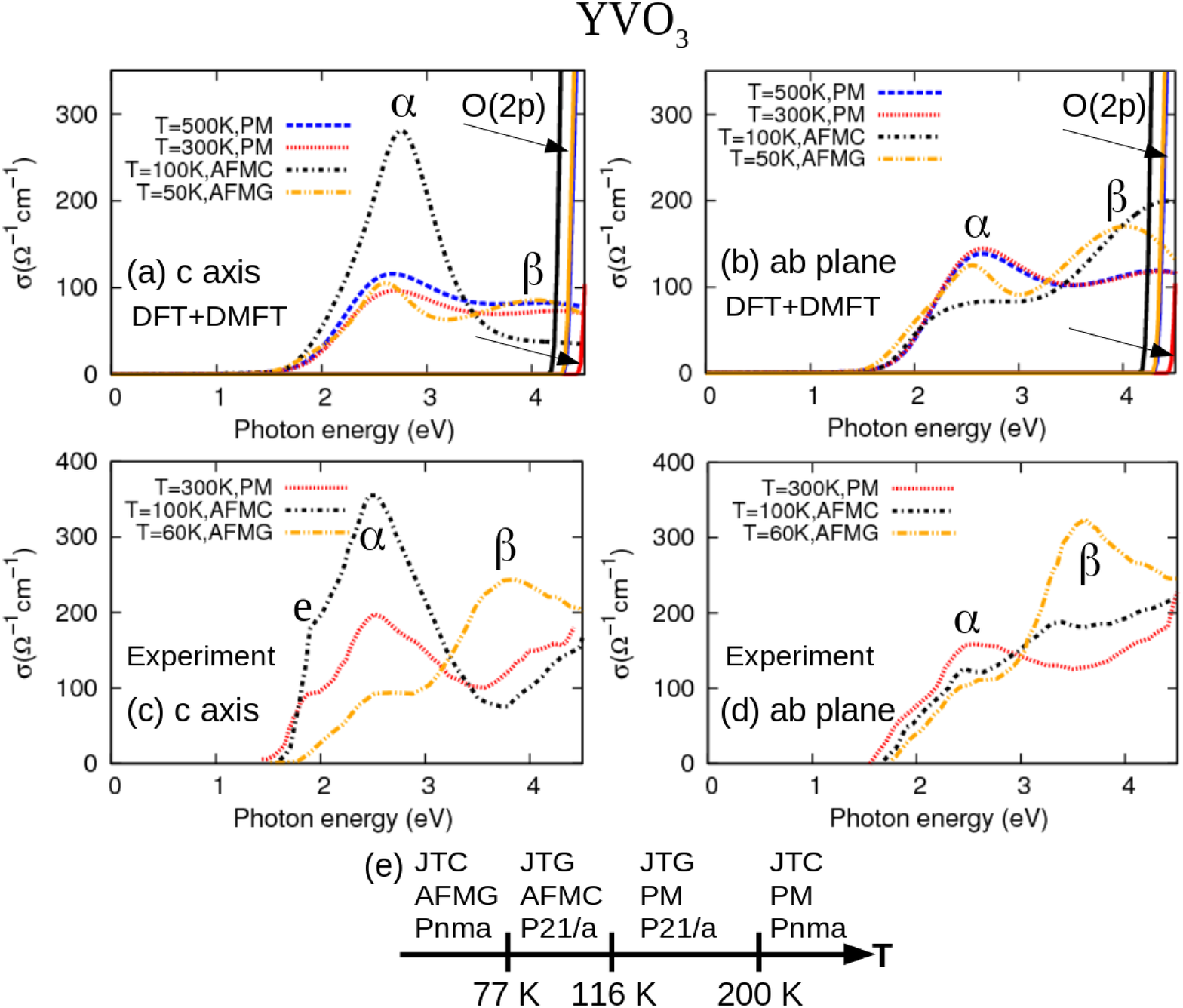}
\caption{(Color)
(a) and (b) Temperature ($T$) dependent optical conductivity $\sigma(\omega)$
of YVO$_{3}$ for $c$ axis and $ab$ plane in DFT+DMFT.
(c) and (d) $T$-dependent $\sigma(\omega)$
of YVO$_{3}$ for $c$ axis and $ab$ plane of experiment, Reul et al., Ref.\onlinecite{Reul_PRB_2012}.
Solid lines in (a) and (b) are
for the edge of O(2$p$)-V(3$d$) optical transitions for each $T$.
The peak, $e$, in (c) is the exciton peak of experiment,
Reul et al., Ref.\onlinecite{Reul_PRB_2012} (see the text).
(e) $T$-dependent phase diagram of YVO$_{3}$.
JTC and JTG type distortions are dominant
in the $Pnma$ and the $P2_{1}/a$ structures, respectively.
PM corresponds to paramagnetism.
\label{fig:OPT_YVO}
}
\end{figure*}
%%%%%%%%%%%%%%%%%%%%%%%%%%%%%%%%%%%%%%%%%%%%%%%%%%%%%%%%%%%%%%%%
%%%%%%%%%%%%%%%%%%%%%%%%%%%%%%%%%%%%%%%%%%%%%%%%%%%%%%%%%%%%%%%%
\begin{figure*}[t]
\includegraphics[width=10cm]{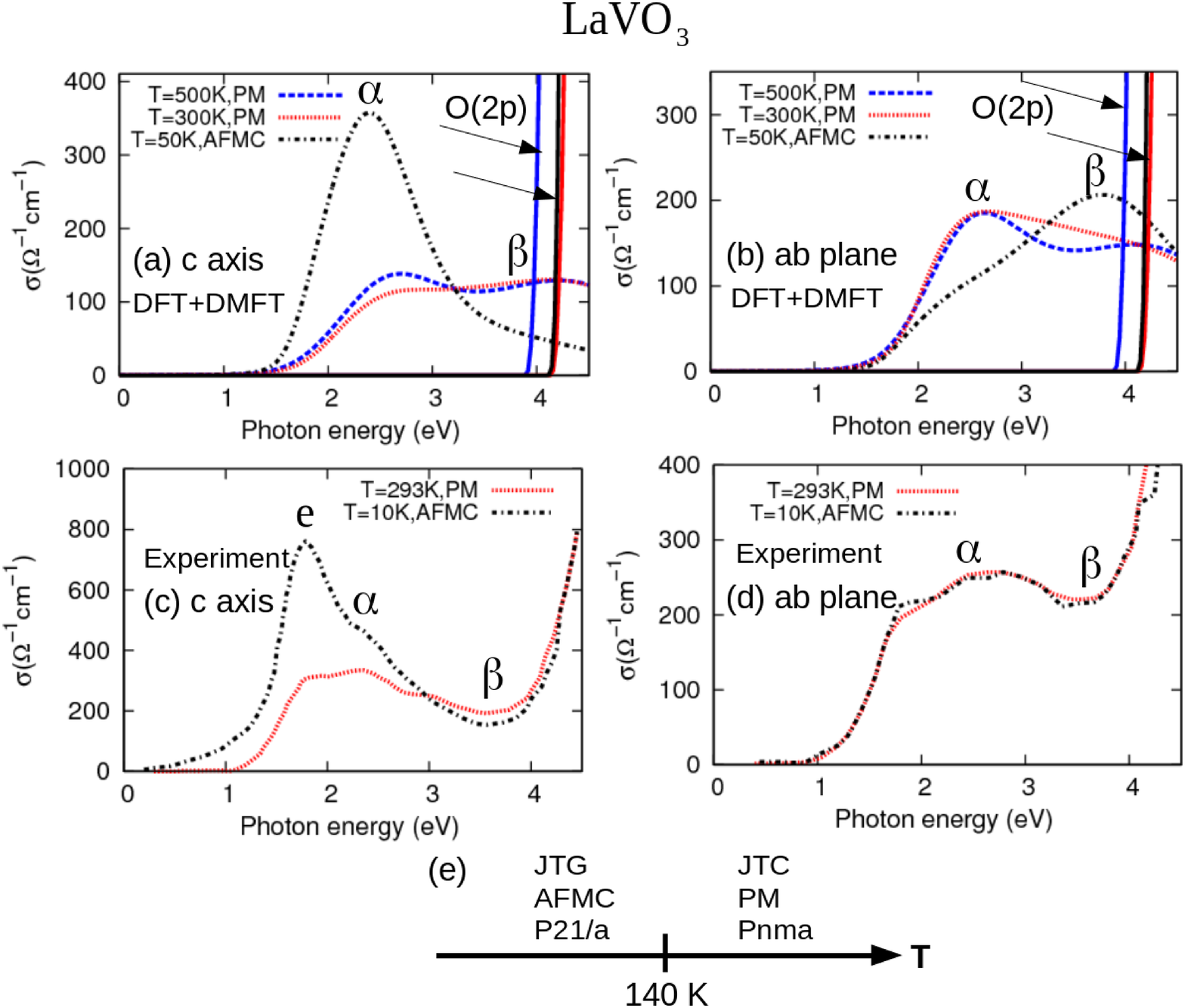}
\caption{(Color)
(a) and (b) $T$-dependent $\sigma(\omega)$
of LaVO$_{3}$ for $c$ axis and $ab$ plane in DFT+DMFT.
(c) and (d) $T$-dependent $\sigma(\omega)$
of LaVO$_{3}$ for $c$ axis and $ab$ plane of experiment, Miyasaka et al., Ref.\onlinecite{Miyasaka_JSPJ_2002}.
Solid lines in (a) and (b) are
for the edge of O(2$p$)-V(3$d$) optical transitions for each $T$.
The peak $e$ in the experiment, in (c) is interpreted as an exciton peak (see the text).
(e) $T$-dependent phase diagram of LaVO$_{3}$.
\label{fig:OPT_LVO}
}
\end{figure*}
%%%%%%%%%%%%%%%%%%%%%%%%%%%%%%%%%%%%%%%%%%%%%%%%%%%%%%%%%%%%%%%%

\section{Methods}
\label{sec:methods}
We calculate the optical conductivity
of YVO$_{3}$ and LaVO$_{3}$ within DFT+DMFT using the
full potential of Ref\cite{aichhorn2009dynamical}
and the TRIQS library.\cite{TRIQS_Package_2015,aichhorn2016triqs}
We compute optical conductivity using
the Kubo formula with the bubble diagram of the Green's function
in the DFT+DMFT.\cite{DMFT_rmp_1996,kotliar2006electronic}
We adopted experimental crystal structures of each relevant $T$
for both YVO$_{3}$ and LaVO$_{3}$.\cite{bordet1993structural,blake2002neutron}
For YVO$_{3}$, (i) for high $T$ (300-500 K), the experimental crystal
structure for $T$=297 K is used ($Pnma$), (ii) for intermediate $T$ (100K),
the experimental crystal structure for $T$=100 K is used ($P2_{1}/a$),
and (iii) for low $T$ (50K),
the experimental crystal structure for $T$=65 K is used ($Pnma$).\cite{blake2002neutron}
For LaVO$_{3}$, (i) for high $T$ (150-500 K), the experimental crystal
structure for $T$=150 K is used ($Pnma$), and
(ii) for low $T$ (50-100K), the experimental crystal structure
for $T$=100 K is used ($P2_{1}/a$).\cite{bordet1993structural}
In the DFT part of the computation, the Wien2k package
was used.\cite{blaha2001wien2k}
The local density approximation (LDA) is used for the exchange-correlation functional.
For projectors on the correlated $t_{2g}$ orbitals in DFT+DMFT, Wannier-like $t_{2g}$ orbitals
are constructed out of Kohn-Sham bands within the energy window [-1.1,1.1] eV with
respect to the Fermi energy.
We use the full rotationally invariant Kanamori
interaction as shown below,
where $L$ and $S$ are angular momentum and
spin momentum operators of $t_{2g}$ orbitals,
$H_{\rm int}=(U-3\J)\frac{\hN(\hN-1)}{2}-2\J\vS^{2}-\frac{\J}{2}\vL^2$.
(see e.g. the algebra in Ref.\onlinecite{georges_Hund_review_annrev_2013})
For $U$ and $J$ parameters of the Kanamori interaction,
we used $U$=4.5 eV and $J$=0.5 eV.
This parameter range is shown in Ref.\onlinecite{de2007orbital}
to be relevant for the description of
experimental photoemission spectrum (Ref.\onlinecite{pen1999electronic,maiti2000spectroscopic}) of YVO$_{3}$ and LaVO$_{3}$ from
the $t_{2g}$ low energy effective model of the DFT+DMFT.\cite{casula2012low}
One thing should be noticed is that in Ref.\onlinecite{de2007orbital},
only the density-density type interaction in the Kanamori interaction is
used. Using
the full rotationally invariant Kanamori interaction including
the spin-flip and the pair-hopping terms is
essential to describe the correct splitting of $\alpha$ and $\beta$ peaks
in the optical conductivity which have different multiplets
in optical transitions as shown in Fig.\ref{fig:Introduction}(d and e).
To solve the quantum impurity problem in the DMFT,
we used the strong-coupling continuous-time Monte Carlo
impurity solver\cite{gull_CTQMC_rmp_2011} as implemented in the TRIQS library.\cite{TRIQS_Package_2015,TRIQS_CTQMC_2016}

\section{Results}
\label{sec:results}

In this section, we present our results for the optical
conductivity of YVO$_{3}$ and LaVO$_{3}$.
In Sec.\ref{subsec:Multiplet},
we assigned main features of optical conductivity,
$\alpha$, $\beta$, and O(2$p$)-V(3$d$) peaks, to the multiplet of
final states of optical transitions, and
also discussed their relation to magnetic states.
In Sec.\ref{subsec:Tdep},
we presented the $T$-dependent evolution of optical conductivity
and compared the result with experiments.
In Sec.\ref{subsec:orbital_state}, we presented
the $T$-dependent the evolution of orbital states in the
$xz$, $yz$, and $xy$ states.
This $T$-dependent orbital state is also computed by imposing PM state.
Under this PM state, the $T$-dependent optical conductivity is discussed.
In Sec.\ref{subsec:LVO145K},
we presented evolution
optical conductivity of LaVO$_{3}$ as function of $T$ around 150K for
various spin states, and suggest possible fingerprints
of the spin state in the optical conductivity for this $T$ regime.

\subsection{Multiplet states and optical conductivity}
\label{subsec:Multiplet}

Two main peaks, $\alpha$ and $\beta$, in the
optical conductivity can be assigned to the
optical transitions having
multiplet configurations in Figure\ref{fig:Introduction}(d and e).
The $\alpha$ peak corresponds to the inter-site electron transfer
within the FM spin states, and the $\beta$ peak corresponds to
the inter-site electron transfer within the AFM spin states.
Figure\ref{fig:PDOS} presents schematic partial density of states (PDOS)
of O(2$p$) and V(3$d$) bands and its interpretation
for the optical transitions, $\alpha$, $\beta$, and O(2$p$)-V(3$d$).
Photon energies for $\alpha$ and $\beta$ peaks
are $U-3J$ and $U$, respectively.
Due to the incomplete orbital polarization of $xz/yz$ and
the rotation of octahedron which breaks cubic symmetry,
the inter-orbital optical transition is also possible.
The spectral weight from the optical transition,
which has a double occupancy in the same orbital
for the final state,
has the corresponding photon
energy of $U+2J$.
This spectral weight is buried in V(3$d$)-O(2$p$)
optical transition as shown in Fig.\ref{fig:PDOS}.
Fig.\ref{fig:OPT_YVO} shows that
the peak position of $\alpha$ and $\beta$
in our DFT+DMFT results (Fig.\ref{fig:OPT_YVO}~a and b)
is consistent with experiments (Fig.\ref{fig:OPT_YVO}~c and d) of Ref.\onlinecite{Reul_PRB_2012}.
The splitting between $\alpha$ and $\beta$ peaks, $3J$($\sim1.5$ eV),
in DFT+DMFT results is consistent with experiments.
This result indicates that the above description of $\alpha$ and $\beta$ peaks is correct.

\subsection{Temperature dependent evolution of optical conductivity}
\label{subsec:Tdep}

Figure~\ref{fig:OPT_YVO} presents the $T$-dependent evolution of the optical
conductivity of YVO$_{3}$.
From the change of $T$ from 500 K to 300 K, in the DFT+DMFT result,
it is shown that there are little changes in the optical
conductivity in both $ab$ plane and $c$ axis.
In this $T$ range, the crystal structure is $Pnma$ and
the spin state is PM.
From the change of $T$ from 300 K to 100 K, there is
a structural transition from $Pnma$ to $P2_{1}/a$,
and accordingly, there is an onset of the AFMC ordering
which is related to the transition of octahedron distortion
from JTC to JTG.
As shown in Fig.\ref{fig:Introduction}(f), this AFMC
ordering induces an enhancement (suppression) of $\alpha$ ($\beta$) peak
for $c$ axis and the opposite trend for $ab$ plane.
In Fig.\ref{fig:OPT_YVO}, it is shown that
the DFT+DMFT result is remarkably consistent with
experiment.\cite{Reul_PRB_2012}
With the change of $T$, from 100 K to $\sim$50 K,
there is a structural transition from $P2_{1}/a$ to $Pnma$
with the onset of the AFMG ordering and
the transition of distortion from JTG to JTC.
In this state, all inter-site spin configurations
are AFM as shown in Fig\ref{fig:Introduction}(g).
As a result, the optical conductivity becomes more isotropic,
with a suppression of $\alpha$ peak and an enhancement of $\beta$ peak
for both $c$ axis and $ab$ plane, with respect to PM state.
This result in DFT+DMFT for 100 K to $\sim$50 K
is consistent with the experiment.\cite{Reul_PRB_2012}
The overall evolution of $\alpha$ peak heights in $c$ axis
in the present DFT+DMFT result, (97, 281, and 105 $\Omega^{-1}cm^{-1}$
for 300, 100, and 50 K, respectively)
is consistent with the experiment (198, 355, and 92 $\Omega^{-1}cm^{-1}$
for 300, 100, and 60 K, respectively).\cite{Reul_PRB_2012}
In the photon energy above 4 eV, the optical transition
of O(2$p$)-V(3$d$) is activated. As a result, 
as shown in Fig.\ref{fig:OPT_YVO}, there is
an upturn of optical conductivities, and
the $\beta$ peak is partially buried, consistent with experiments.\cite{Reul_PRB_2012}

%%%%%%%%%%%%%%%%%%%%%%%%%%%%%%%%%%%%%%%%%%%%%%%%%%%%%%%%%%%%%%%%
\begin{figure*}[t]
\includegraphics[width=10cm]{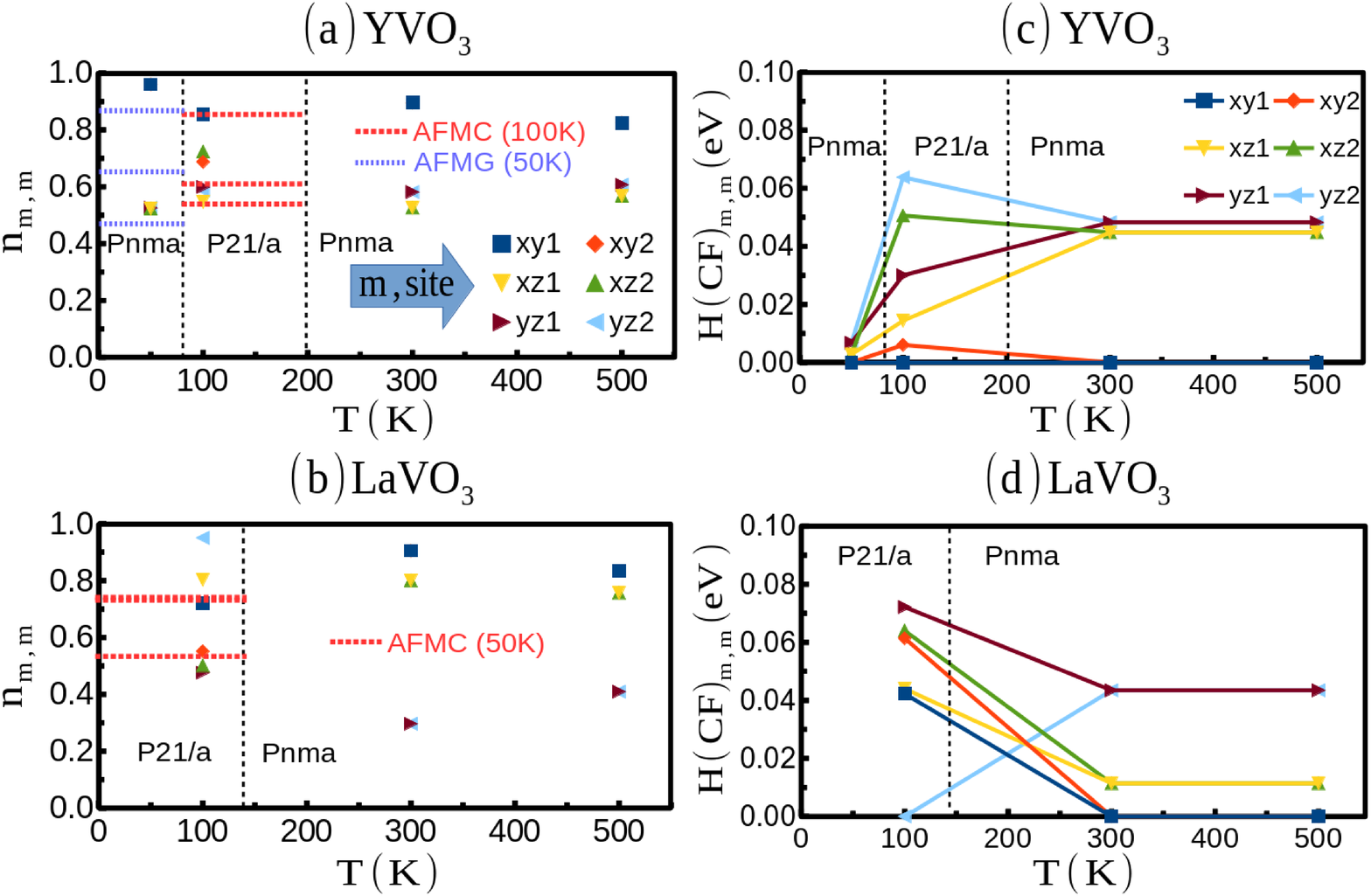}
\caption{(Color) (a) and (b)
$T$-dependent orbital density of YVO$_{3}$ and LaVO$_{3}$, respectively.
In the case of the $P2_{1}/a$ space group, two sites
,1 and 2, are symmetrically non-equivalent.
Dots are orbital densities in the PM states.
Red (dashed) and blue (dotted) lines are orbital densities
in the AFMC and AFMG states in Fig.\ref{fig:OPT_YVO} and Fig.\ref{fig:OPT_LVO},
which are averaged per site. The deviation from the averaged value
in AFM states is less than 0.077.
(c) and (d) $T$ dependent evolution of crystal-field
for YVO$_{3}$ and LaVO$_{3}$, respectively, 
for crystal structures used in the present DFT+DMFT computation.\cite{blake2002neutron,bordet1993structural} 
The lowest crystal-field
energy level is set as a zero energy for each structures.
\label{fig:Orbital_State}
}
\end{figure*}
%%%%%%%%%%%%%%%%%%%%%%%%%%%%%%%%%%%%%%%%%%%%%%%%%%%%%%%%%%%%%%%%
%%%%%%%%%%%%%%%%%%%%%%%%%%%%%%%%%%%%%%%%%%%%%%%%%%%%%%%%%%%%%%%%
\begin{figure}[t]
\includegraphics[width=\columnwidth]{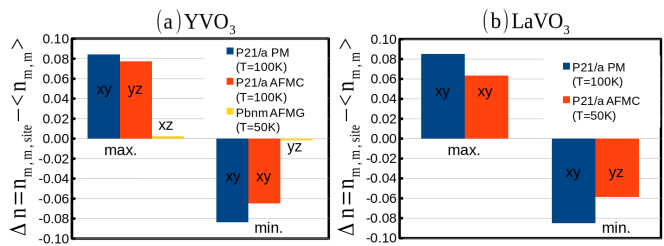}
\caption{(Color) (a) The maximum and the minimum
of orbital density deviations within all 
non-equivalent site from the site averaged orbital density,
for each spin state for YVO$_{3}$.
Orbital which has the maximum or the minimum
are presented for each cases.
(b) Same as (a) for LaVO$_{3}$. It is shown that the
onset of magnetism results in the reduction of the deviation
for both YVO$_{3}$ and LaVO$_{3}$ for $P2_{1}/a$ structure.
\label{fig:Orbital_spin_relation}
}
\end{figure}
%%%%%%%%%%%%%%%%%%%%%%%%%%%%%%%%%%%%%%%%%%%%%%%%%%%%%%%%%%%%%%%%
%%%%%%%%%%%%%%%%%%%%%%%%%%%%%%%%%%%%%%%%%%%%%%%%%%%%%%%%%%%%%%%%
\begin{figure*}[t]
\includegraphics[width=10cm]{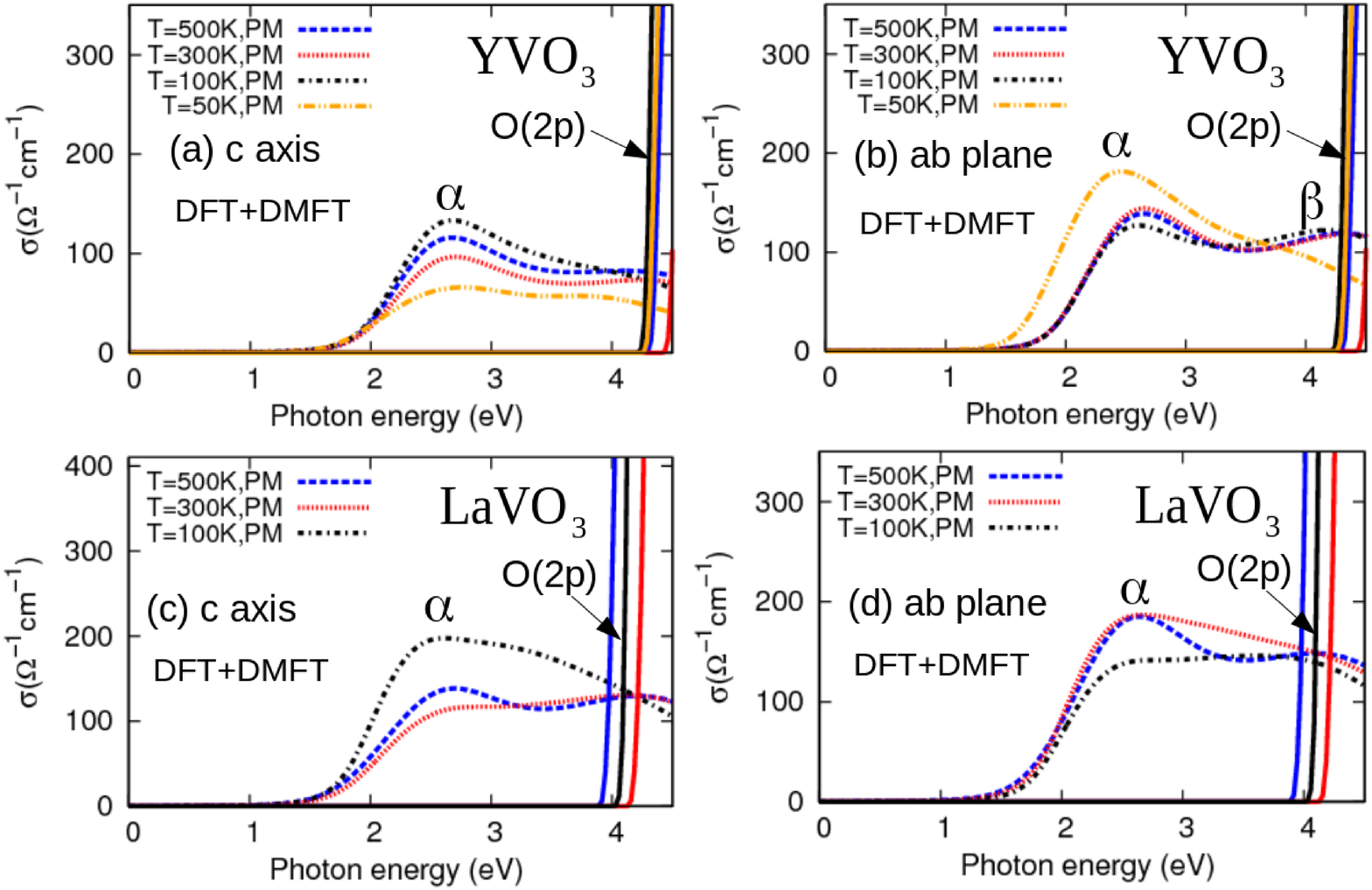}
\caption{(Color)
\label{fig:OPT_PM}
(a) and (b) $T$-dependent $\sigma(\omega)$
of YVO$_{3}$ for $c$ axis and $ab$ plane in DFT+DMFT with
the constraint of PM state.
(c) and (d) $T$-dependent $\sigma(\omega)$
of LaVO$_{3}$ for $c$ axis and $ab$ plane in DFT+DMFT with
the constraint of PM state.
Solid lines are
for the edge of O(2$p$)-V(3$d$) optical transitions.
}
\end{figure*}
%%%%%%%%%%%%%%%%%%%%%%%%%%%%%%%%%%%%%%%%%%%%%%%%%%%%%%%%%%%%%%%%

There are several features in the experimental optical conductivity which are not resolved in our DFT+DMFT computation.
There is a sub-peak, $e$, below the $\alpha$ peak in the experiment (Fig.\ref{fig:OPT_YVO}(c)).
This peak is the exciton below the band gap as shown in Ref.\onlinecite{Reul_PRB_2012}.
The non-equilibrium evolution of the $e$ peak and the $\alpha$ peak
in the pump probe experiment of YVO$_{3}$ in Ref.\onlinecite{novelli2012ultrafast}
shows that this subpeak below the $\alpha$ is indeed the exciton below the band gap.
This assignment of peaks is different from previous reports in $R$VO$_{3}$
of Ref.\onlinecite{Miyasaka_JSPJ_2002,fang2003anisotropic},
which argue that the $e$ and the $\alpha$ peaks
in our viewpoint are $\alpha$ and $\beta$ peaks 
in Fig.\ref{fig:Introduction}(d and e) and Fig.\ref{fig:PDOS}, respectively.
This is the reason why previous optical conductivity data from DFT
is different from experiments.\cite{fang2003anisotropic}
The emergence of the $e$ peak is from the condensation of the exciton below the band gap
which is induced by non-local correlations, 
interpreted as a condensation of the charge carrier state along the ferromagnetic chain in the AFMC state.\cite{Reul_PRB_2012}
Present DFT+DMFT computation includes local dynamical
electronic correlations by the non-perturbative manner and
spatial static electronic correlations from the exchange-correlation
functional in the LDA. As a result, the $e$ peak from the
non-local correlation is not described in the present results.

Figure \ref{fig:OPT_LVO} presents the $T$-dependent
evolution of optical conductivity for LaVO$_{3}$.
For $T$ range from 500 K to 300 K, in the DFT+DMFT result,
it is shown that there is a small change in the optical
conductivity in both $ab$ plane and $c$ axis, similar to
the case of YVO$_{3}$.
With changes of $T$ from 300 K to 50 K,
the structural transition from $Pnma$ to $P2_{1}/a$ occurs.
Accordingly, there is an onset of the AFMC ordering
which is related to the transition from JTC to JTG.
Similar to the case of YVO$_{3}$ for 300 K to 100 K,
this AFMC ordering induces an enhancement
of $\alpha$ peak heights
for $c$ axis and the opposite trend in the $ab$ plane.
As shown in Fig.\ref{fig:OPT_LVO}, the DFT+DMFT
result for 300 K to 50 K is consistent with the experiment for $c$ axis.\cite{Miyasaka_JSPJ_2002}
We suggest that the $e$ peak in Fig.\ref{fig:OPT_LVO}(c) is the exciton
peak similar to the YVO$_{3}$.
For the optical
conductivity of $ab$ plane, differently from the DFT+DMFT result,
there is a small evolution of optical conductivity
in the experiment, which is remained to be resolved.\cite{Miyasaka_JSPJ_2002}
The overall evolution of $\alpha$ peak heights in $c$ axis
in the present DFT+DMFT result, (120, and 357 $\Omega^{-1}cm^{-1}$
for 300, and 50 K, respectively)
is consistent with the experiment (330, and 570 $\Omega^{-1}cm^{-1}$
for 293, and 10 K, respectively).\cite{Miyasaka_JSPJ_2002}

There are two important differences in the optical conductivity for YVO$_{3}$ and LaVO$_{3}$.
Firstly, due to the smaller rotation of the octahedron
in LaVO$_{3}$, which give rise to the enhanced O(2$p$)-V(3$d$) hybridization,
the optical transition of O(2$p$)-V(3$d$) is activated
at the lower photon energy (around 4 eV) in LaVO$_{3}$ with respect to that in YVO$_{3}$ (above 4 eV).
As a result, as shown in Fig.\ref{fig:OPT_LVO}, 
the $\beta$ peak center is buried from the O(2$p$)-V(3$d$) peak in LaVO$_{3}$,
consistent with experiments.\cite{Miyasaka_JSPJ_2002}
This difference of the energy of the O(2$p$)-V(3$d$) transition explains
the presence of the $T$ dependent evolution of the $\beta$ peak in YVO$_{3}$ and
the absence of the $T$ dependent evolution of the $\beta$ peak in LaVO$_{3}$.\cite{Reul_PRB_2012,Miyasaka_JSPJ_2002}
Therefore, we suggest to analyse (i) $T$-dependent evolution of the $\alpha$ peak
to resolve spin-orbital structure in the case of bulk LaVO$_{3}$ and
(ii) the appearance of the $\beta$ peak in the heterostructure of LaVO$_{3}$
in the case of compressive strain to confirm the 
reduced O(2$p$)-V(3$d$) hybridization from the enhanced rotation of octahedron.
Secondly, there is a difference in the hight of the $e$ peaks.
The height of $e$ peaks are 179 and 760 $\Omega^{-1}cm^{-1}$
for YVO$_{3}$ and LaVO$_{3}$, respectively.
We suggest that this larger height of the $e$ peak in LaVO$_{3}$
is due to the smaller octahedron rotation
which results in the larger stabilization
of the kinetic energy of the exciton for LaVO$_{3}$.

\subsection{Orbital states dependence of optical conductivity}
\label{subsec:orbital_state}

Figure~\ref{fig:Orbital_State} presents the $T$-dependent
evolution of orbital density and crystal-field
of $xz$, $yz$, and $xy$ orbital in the $t_{2g}$ manifold
for YVO$_{3}$ and LaVO$_{3}$.
None of orbitals are fully polarized
due to the low symmetry crystal-field contribution
from the rotation of the octahedrons and
the covalent bonding with cations.\cite{de2007orbital}

For $T$ range of 300-500 K, in Fig.\ref{fig:Orbital_State}(a) and
\ref{fig:Orbital_State}(b),
there is a small evolution of orbital polarization
which is consistent with Ref.\onlinecite{de2007orbital}
This result explains the small evolution of the
optical conductivity in the $T$ range of 300-500 K,
as shown in Fig.\ref{fig:OPT_YVO} and \ref{fig:OPT_LVO}.
Fig.\ref{fig:Orbital_State}(c) and (d) 
shows $T$ dependent crystal-field levels in the 
crystal structures used in present results.
The $Pnma$ structure has a single type of crystal-field level 
for all sub-lattice of V (site-uniform), 
and the $P2_{1}/a$ structure has two types 
of crystal-field level for sub-lattice of V (site-non-uniform).
In the PM state of this $T$ range,
the site-uniform crystal-field splitting,
is not effective to induce the $T$-dependent evolution of orbital polarization.

Figure \ref{fig:Orbital_spin_relation}
shows that in the AFMC state,
the orbital polarization of $xz$ and $yz$ is more uniform
than that of the PM state for $P2_{1}/a$ structure. Also, 
as shown in Fig.\ref{fig:Orbital_State}(a), especially for
the $Pnma$ structure of YVO$_{3}$ for $T$=50 K, the orbital
polarization in the AFMG state is much larger than that of the PM state.
These results suggest that JT distortions
from the structural transition induce a finite orbital polarization,
and the superexchange interaction modifies this orbital polarization
with the onset of AFM spin ordering.
The size of the induced magnetic moment for each site is 1.97 $\mu_{B}$ for AFMG state.
This large magnetic moment induced exchange field 
to orbitals results in the large orbital polarization in the AFMG state 
with respect to that in the PM state as shown in Fig.\ref{fig:Orbital_State}(a).
This result shows that both JT distortion and
superexchange interaction contribute to
the orbital polarization.
This result is in line with the inelastic x-ray scattering experiment
on YVO$_{3}$ which shows that the orbital excitation
has contributions from
both superexchange and crystal-field.\cite{benckiser2013orbital}

Fig. \ref{fig:Orbital_spin_relation} shows
the maximum and the minimum
of orbital density deviations within all site
from the site averaged orbital density, 
for each spin states, for YVO$_{3}$ and LaVO$_{3}$.
It is shown that the onset of magnetic ordering
results in the reduction of the site dependence of the orbital polarization
in both YVO$_{3}$ (Fig.\ref{fig:Orbital_spin_relation}(a))
and LaVO$_{3}$ (Fig.\ref{fig:Orbital_spin_relation}(b)).
As shown in Fig.\ref{fig:Orbital_State}(c) and (d),
the crystal-field in the structure of $P2_{1}/a$ space group
is non-uniform for sites. On the other hand, the magnetic moment
for magnetic states of all case is uniform for sites.
The site dependent deviation of
magnetic moment size is smaller than 0.01 $\mu_{B}$.
As a result, the site-uniform orbital polarization from
the site-uniform spin exchange interaction is induced
in all AFM state. 
This site-uniform orbital state in AFM states 
is different from the crystal-field driven site-non-uniform orbital polarization, 
which is induced for PM states of the $P2_{1}/a$ space group.

In $P2_{1}/a$ structures for YVO$_{3}$ and LaVO$_{3}$,
the orbital polarization for $xz$ and $yz$ is larger
in the case of LaVO$_{3}$ for PM spin state.
This result is due to the nature of the JT distortion
in the $P2_{1}/a$ structure.
In Ref.\onlinecite{Varignon_electrical_control},
it was shown that both JTC and JTG distortions
exist for $P2_{1}/a$ structures of YVO$_{3}$ and LaVO$_{3}$.
In the case of LaVO$_{3}$,
JTG type distortion is much larger than JTC type distortion.
On the one hand, YVO$_{3}$ has also larger JTG type distortion
than that of JTC, but the difference is smaller than that of LaVO$_{3}$.\cite{Varignon_electrical_control}
As a result, the orbital polarization is
larger for LaVO$_{3}$ with respect to that of YVO$_{3}$ for $P2_{1}/a$ structure in PM state,
consistent with Ref.\onlinecite{de2007orbital}, as shown in Fig.\ref{fig:Orbital_State}(a and b).

Figure~\ref{fig:OPT_PM} presents the
$T$-dependent evolution of optical conductivity in
DFT+DMFT with
the constraint of PM state.
With the constraint of PM state, Kramers doublet
,spin up and spin down components, are set to equal,
and orbital differentiation is allowed with constraint 
from the given symmetry of crystal structure.
Even in the constraint of PM,
due to the superexchange energy gain according to the
GKA rules, antiferro OO and ferro OO
enhances FM ($\alpha$ peak) and AFM ($\beta$ peak)
spin correlations, respectively.
As a result, the orbital polarization in the PM state also
contributes to the evolution of optical conductivity.
For the structural transition from the $Pnma$ to the $P2_{1}/a$,
300K to 100K of $T$, LaVO$_{3}$ shows a large evolution of
optical conductivity, while YVO$_{3}$ shows a small variation.
The larger JTG type orbital polarization in the $P2_{1}/a$
structure of LaVO$_{3}$ induces the AFMC type spin correlation.
On the other hand, the optical conductivity
of YVO$_{3}$ for $P2_{1}/a$ structure shows small
variations because of the small difference in the
magnitude of JTG and JTC type distortions.
Thus, the trend, such that $\alpha$ ($\beta$) peak
is much enhanced (suppressed) for $c$ axis
and the opposite change occurs for $ab$ plane,
is larger for LaVO$_{3}$.
This trend is consistent with the larger orbital polarization
in LaVO$_{3}$ with respect to that in YVO$_{3}$ for $P2_{1}/a$
structures in PM state as shown in Fig.\ref{fig:Orbital_State} (a) and (b).

%%%%%%%%%%%%%%%%%%%%%%%%%%%%%%%%%%%%%%%%%%%%%%%%%%%%%%%%%%%%%%%%
\begin{figure}[b]
\includegraphics[width=\columnwidth]{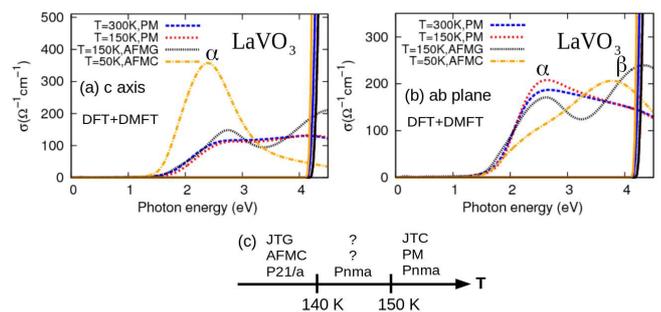}
\caption{(Color)
(a) and (b) $T$-dependent $\sigma(\omega)$
of LaVO$_{3}$ for $c$ axis and $ab$ plane in DFT+DMFT
including $\sigma(\omega)$ at $T$=150 K ($Pnma$ space group)
for PM and AFMG states.
The AFMC state is not stabilized at $T$=150 K.
(c) $T$-dependent phase diagram of LaVO$_{3}$ around 150 K.
Solid lines are
for the edge of O(2$p$)-V(3$d$) optical transitions.
\label{fig:OPT_LVO_150K}
}
\end{figure}
%%%%%%%%%%%%%%%%%%%%%%%%%%%%%%%%%%%%%%%%%%%%%%%%%%%%%%%%%%%%%%%%

For YVO$_{3}$, in Fig.\ref{fig:Orbital_State}(a),
it is shown that in the $Pnma$ structure of $T$=50K for PM state,
the orbital occupancy of the $xy$ orbital is close to 1.
And also, there is unquenched orbital fluctuation of $xz$ and $yz$ for the PM state.
This result provides an explanation of the optical conductivities 
of $T$=50K for PM state in Fig.\ref{fig:OPT_PM}(a and b).
With the constraint of
ferro OO in $c$ axis (OOC) from the JTC distortion of $Pnma$ structure,
the orbital state of $xz$ and $yz$ in the apical direction
has same phase. As a result, there is a strong suppression of optical transition for $c$ axis
due to the blocking of the charge transfer for $c$ axis.
On the other hand, the orbital fluctuation of $xz$ and $yz$ 
is in different phase for $ab$ plane due to the symmetry of $Pnma$,
which enhances in-plane FM correlation. 
As a result, the height of $\alpha$ peak
is enhanced for $ab$ plane.
These results suggest that the optical conductivity is
a useful quantity to probe the orbital state of $R$VO$_{3}$,
which is complimentary to the resonant x-ray diffraction experiment.
We suggest to compare heights of the $\alpha$ peak for
consistent analysis of spin-orbital states from optical conductivity
of $R$VO$_{3}$ systems, because, in the case of LaVO$_{3}$, the $\beta$ peak
is buried in the O(2$p$)-V(3$d$) peak.

\subsection{Spin state and optical conductivity of LaVO$_{3}$ for T$_{str}$<T<T$_{N}$ (T$\sim$150K)}
\label{subsec:LVO145K}

Figure~\ref{fig:OPT_LVO_150K} presents
$T$-dependent evolution of optical conductivity of LaVO$_{3}$
in the DFT+DMFT computation
at the $T$ above the structural transition ($T_{str}$)
and below the $T$ for the spin ordering ($T_{N}$)
,which corresponds to $T\sim$150 K.
At this $T$, there is a debate on the spin ordering.\cite{khaliullin2005orbital,zhou2008frustrated}
The structural distortion of JTC in the $Pnma$ space group
above $T_{str}$ induces the AFMG ordering.
On the other hand, the superexchange driven orbital RVB state induces AFMC.
We compute the optical conductivity of
AFMG and PM states at 150 K in the structure of the $Pnma$ space group.
For this $T$, the AFMC state is not stabilized,
which implies that the description of the free energy
in the present level of approximation in the DFT+DMFT
does not have a local minimum for the AFMC state for the
structure of $T$=150 K.
The result is compared with results for the AFMC state
at 100 K ($P2_{1}/a$ structure) and the PM state at 300 K ($Pnma$ structure).
It is clearly shown that in the case that the AFMG ordering
emerges for $T_{str}$<$T$<$T_{N}$, the $\alpha$ peak is
suppressed for $c$ axis with respect to the case for the AFMC state.
This trend from the AFMG state is apparently different from
the anisotropic optical conductivity for the AFMC state.
We suggest that the measurement of the evolution of optical conductivity
in the regime of $T_{str}$<$T$<$T_{N}$ would be useful for resolving
the issue on spin ordering between the JTC structure driven AFMG and
the electronic superexchange induced orbital RVB driven AFMC.

\section{Discussion}
\label{sec:discussion}
We have shown that the $T$-depedent evolution of optical conductivty
of YVO$_{3}$ and LaVO$_{3}$ has signatures of spin and orbital states
in these materials.
DFT+DMFT reproduces experimental results
from the correct assignment of multiplet states
to peaks in the optical conductivity.
Two types of magnetic state, AFMG and AFMC, could be resolved
by the optical conductivity.
Furthermore, we have shown that
there is a fingerprint of the $T$-dependent evolution of the orbital polarization
in the optical conductivity even in the absence of the spin order.
This result provides a reference to probe orbital states
in the heterostructure of $R$VO$_{3}$.\cite{weng2010phase,sclauzero2015structural}
The clear difference between AFMG and AFMC states indicates that
the optical conductivity is useful for resolving
issues on the spin ordering of LaVO$_{3}$ for the regime of $T_{str}$<$T$<$T_{N}$,
which depends on its origin between
the electronic superexchange induced orbital RVB and the JT crystal-field induced OO.

\acknowledgements
We acknowledge useful discussions with Antoine Georges, Stefano Gariglio, Hugo Meley,
and Alaska Subedi.
This work was supported by the European Research Council (ERC-319286 QMAC),
and by the Swiss National Science Foundation (NCCR MARVEL).

\bibliography{refsRVO}

\end{document}